\documentclass[twocolumn,aps,prl,showpacs]{revtex4-1}
\usepackage{amssymb,amsmath,amsfonts}
\usepackage{epsfig,graphicx}
\usepackage{hyperref}
\hypersetup{colorlinks=true,citecolor=NavyBlue,filecolor=black,linkcolor=black,urlcolor=black}
\usepackage{color}
\usepackage[dvipsnames]{xcolor}
\definecolor{blueRP}{rgb}{0.0, 0.58, 0.71}
\definecolor{purpleMP}{rgb}{0.6, 0, 0.7}

\newcommand{\adag}{a^{\dagger}}
\newcommand{\adaga}{a^{\dagger}a}
\newcommand\ket[1]{\left|#1\right\rangle}
\newcommand\bra[1]{\left\langle #1 \right|}

\newcommand\HJC{H_{\rm JC}}

\begin{document}

\date{\today}
\title[]{Quantum state engineering by shortcuts-to-adiabaticity in interacting spin-boson systems}
\author{Obinna Abah}
\thanks{O. A. and R. P. contributed equally to this work.}\email{o.abah@qub.ac.uk}
\affiliation{Centre for Theoretical Atomic, Molecular and Optical Physics, School of Mathematics and Physics, Queen's University Belfast, Belfast BT7 1NN, United Kingdom}
\author{Ricardo Puebla}\email{r.puebla@qub.ac.uk}
\affiliation{Centre for Theoretical Atomic, Molecular and Optical Physics, School of Mathematics and Physics, Queen's University Belfast, Belfast BT7 1NN, United Kingdom}
\author{Mauro Paternostro}
\affiliation{Centre for Theoretical Atomic, Molecular and Optical Physics, School of Mathematics and Physics, Queen's University Belfast, Belfast BT7 1NN, United Kingdom}

\begin{abstract}
  We present  a fast and robust framework to prepare non-classical states of a bosonic mode exploiting a coherent exchange of excitations with a two-level system ruled by a Jaynes-Cummings interaction mechanism. Our protocol, which is built on shortcuts to adiabaticity,  allows for the generation of arbitrary Fock states of the bosonic mode, as well as coherent quantum superpositions of a Schr\"odinger cat-like form. In addition, we show how to obtain a class of photon-shifted states where the vacuum population is removed, a result akin to photon addition, but displaying more non-classicality than standard photon-added states. Owing to the ubiquity of the spin-boson interaction that we consider, our proposal is amenable for implementations in state-of-the-art experiments.
\end{abstract}

\maketitle

{\em Introduction.---} Quantum state engineering, i.e. the manipulation and control of a quantum system to attain a target state with high fidelity, lies at the core of quantum-based technologies~\cite{Nielsen}. In this realm, non-classical states are of key significance to exploit quantum resources and find numerous applications in different areas, such as quantum information processing~\cite{Vogel:93}, sensing~\cite{Degen:17} and fundamental physics inquiries~\cite{Hornberger:12,Bassi:13}. In particular, hybrid quantum systems are well suited to operate as fundamental building block for the engineering of non-classical states and the implementation of the aforementioned tasks~\cite{Jeong:02,Ralph:03,Pan:01,Curty:04,Jennewein:00,McDonnell:07,Brask:10,Sangouard:10,Fischer:15,Schleier-Smith:16}. Such systems can be controlled and manipulated with a very high accuracy in distinct state-of-the-art experiments, such as setups based on trapped ions~\cite{Leibfried:03,Haffner:08}, ensembles of NV centers embedded in a single-crystal diamond nanobeam~\cite{Doherty:12} and superconducting qubits~\cite{Devoret:13}. Since the first generation of a non-classical state, attained in a trapped-ion experiment~\cite{Meekhof:96}, other realizations have been achieved in different experimental setups, e.g.~\cite{Hofheinz:08}. However, the preparation of non-classical states is challenging as they are prone to decoherence, and thus fragile against noise sources. Fast and robust protocols are therefore valuable for their successful preparation, such as the application of stimulated Raman adiabatic passages~\cite{Premaratne:17,Bergmann:98} or dynamical decoupling schemes~\cite{Viola:99,Souza:12,Lidar:14}. Yet, dissipation and distinct noise sources can have a significant impact in their performance, typically requiring a trade-off with slow evolution times.

To circumvent these drawbacks, the current efforts are geared towards the design of protocols at the coherent level, dubbed as shortcuts to adiabaticity (STA), aiming at speeding up the quantum adiabatic process~\cite{Chen2010PRL,Torrontegui:13,Zhou:16} (see Refs.~\cite{MartinezGaraot:15,Palmero:19} for fast quasiadiabatic dynamics protocols). Owing to short evolution times, these protocols are intrinsically resilient to decoherence effects. The counterdiabatic driving requires an additional term that suppresses non-adiabatic transitions between instantaneous eigenstates~\cite{Berry2009JPA}. This active field of research is finding numerous applications in distinct areas, ranging from aspects of many-body physics~\cite{delCampoPRL2012,Campbell:15} to the design of  super-efficient quantum engines~\cite{GooldSciRep,Abah:19pre}, allowing for the design of robust protocols [cf. Ref.~\cite{Odelin:19} for a review].

In this paper, we present a scheme that allows for a fast and robust preparation of non-classical states built on STA and making use of the ubiquitous Jaynes-Cummings (JC) interaction between a two-level system and a single bosonic mode. To illustrate the performance and versatility of the reported protocol, we show how to generate Fock states, Schr\"odinger cat states and strongly non-classical states akin to excitation-added states~\cite{Kim:08}, with very high fidelity and in a short evolution time compared to their adiabatic preparation.
 Finally, we comment on the robustness and noise resilience of our protocol, and its experimental implementation, which is amenable in  state-of-the-art quantum optics setups, such as cavity/circuit quantum electrodynamics, trapped ions and optomechanics. 

{\em General framework.---}
At the heart of quantum optics, the JC model~\cite{Jaynes:63} describes the coupling between light and matter through a simple mechanism connecting a two-level system and a bosonic mode. The relevance of this model goes beyond the scope of pure light-matter interaction and correctly describes spin-phonon coupling, essential for example in trapped ions~\cite{Leibfried:03,Haffner:08} and electromechanical setups~\cite{Treutlein:14}.  The Hamiltonian of this model reads (we choose units such that $\hbar =1$)
\begin{align}\label{eq:HJC}
H_{\rm JC}(t) =\omega_q(t) \sigma_z/2 + \omega \adaga + \lambda(t)(a\sigma^++\adag \sigma^-),
\end{align}
where $\omega_q(t)$ ($\omega$) is the two-level (bosonic) frequency and $\lambda(t)$ is the interaction strength between such subsystems.  The two-level system is characterized by the ladder operators $\sigma^+\!=\!{\sigma^-}^\dag\!=\!\ket{e}\bra{g}$, $\sigma_z\!=\!\ket{e}\bra{e}-\ket{g}\bra{g}$, with $\ket{g}$ and $\ket{e}$ the fundamental and excited state of the two-level system, %with $\sigma_{x}\!=\!\sigma^++\sigma^-$ and $\sigma_{y}\!=\!-i(\sigma^+-\sigma^-)$, 
while the bosonic mode is described by the annihilation and creation operators $a$ and $\adag$ with $[a,\adag]\!=\!1$.
Without loss of generality, we assume that the driving is performed on the frequency of the two-level system $\omega_q(t)$ and the coupling rate $\lambda(t)$, while the bosonic frequency remains constant.
As the total number of excitations $N_e\equiv\ket{e}\bra{e} + \adaga$ is conserved, $H_{\rm JC}(t)$ can be diagonalized in the subspace spanned by $\left\{\ket{e,n},\ket{g,n+1} \right\}$, where $\ket{n}$ ($n\!=\!0,1,\ldots$) is the $n$-excitation Fock state of the mode.  We thus have $H_{\rm JC}(t)\!=\!-\omega_q(t)/2 \ket{g,0}\bra{g,0}+\bigoplus_{n}H_n(t)$ %with $\ket{g,0}$ left uncoupled, and $H_{n}(t)$ is the dressed basis Hamiltonian~\cite{sup}. For convenience, we express 
with the Landau-Zener-like terms $H_{n}(t)={(n+1/2)\omega}\mathbb{I}+({\delta(t)}/{2})\bar{\sigma}_z+\lambda(t)\sqrt{n+1}\bar{\sigma}_x$ and the spin-like operators $\bar{\sigma}^-\!=\!\ket{g,n+1}\bra{e,n}$, $\bar{\sigma}^+\!=\!\ket{e,n}\bra{g,n+1}$, $\bar{\sigma}_z\!=\!\ket{e,n}\bra{e,n}-\ket{g,n+1}\bra{g,n+1}$ ($\delta(t)\!=\!\omega_q(t)-\omega$ is the detuning from atomic resonance).

{\em Shortcut to adiabaticity.---} In general, driving under $H_{\rm JC}(t)$ leads to a  non-adiabatic evolution. Adiabatic evolution is achieved when $\omega_q(t)$ and $\lambda(t)$ vary slowly, i.e. in a time much larger than the typical time scale of the system given by the inverse of the minimum energy gap of $H_{\rm JC}(t)$~\cite{Messiah}. This process can be sped-up by introducing an additional term to the bare Hamiltonian, whose form is given by $H_{\rm CD}(t)\!=\!i \sum_{n,\sigma=\pm} [\partial_t\Phi_{n,\sigma}(t),\Phi_{n,\sigma}(t)]$ with $\Phi_n(t)=\ket{n,\sigma(t)}\bra{n,\sigma(t)}$ and $\ket{n,\sigma(t)}$ denoting the dressed-atom eigenstates of the $H_{\rm JC}(t)$~\cite{Demirplak2003,Berry2009JPA}. 
%H_\text{CD}(t) &= i \sum_{n,\sigma=\pm}  (\ket{\partial_t (n,\sigma(t))} \bra{n,\sigma (t)} \\ &-  \braket{n,\sigma (t)}{\partial_t (n,\sigma(t))} \ket{n,\sigma (t)}\bra{n,\sigma (t)}), 
%with $\ket{n,\sigma (t)}$ denoting the dressed-atom eigenstates of the $H_{\rm JC}(t)$.
%
The resulting  counterdiabatic Hamiltonian reads $H_\text{CD}(t)=i\theta(t)(\adag\sigma^--a\sigma^+)$~\cite{sup} with 
\begin{equation}
%H_\text{CD}(t)= i\frac{\dot{\lambda}(t)\delta(t) - \lambda(t)\dot{\omega}_q(t)}{\delta^2(t)+ \Omega_n^2(t)}(\adag\sigma^--a\sigma^+),
%
\theta(t)\!=\!\frac{\delta(t)\dot{\lambda}(t)-\lambda(t)\dot{\omega}_q(t)}{\Omega_n^2(t)+\delta^2(t)},
\end{equation}
where the parameter $\Omega_n(t)\!=\!2\lambda(t)\sqrt{n+1}$ accounts for a time-varying Rabi frequency in the $n$-subspace.
This additional driving %in the $\bar{\sigma}_y$ direction 
suppresses non-adiabatic excitations allowing for an arbitrarily fast adiabatic evolution. To ensure that the effective Hamiltonian $H_{\rm CD}^{\rm STA}(t)\!=\!\HJC(t) + H_\text{CD}(t)$ equals the original $\HJC(t)$ at the start and end of the protocol, we impose the condition $\dot{\lambda}(0)\!=\!\dot{\lambda}(\tau)\!=\!0$ as well as $\dot{\omega}_q(0)\!=\!\dot{\omega}_q(\tau)\!=\!0$. These conditions ensure $H_{\rm CD}^{\rm STA}(t=0,\tau)\!=\!\HJC(t=0,\tau)$, which can be recast in finding protocols such that $H_{\rm CD}(t=0,\tau)\!=\!0$.

\begin{figure}[t]
\centering
\includegraphics[width=0.9\linewidth,angle=-0]{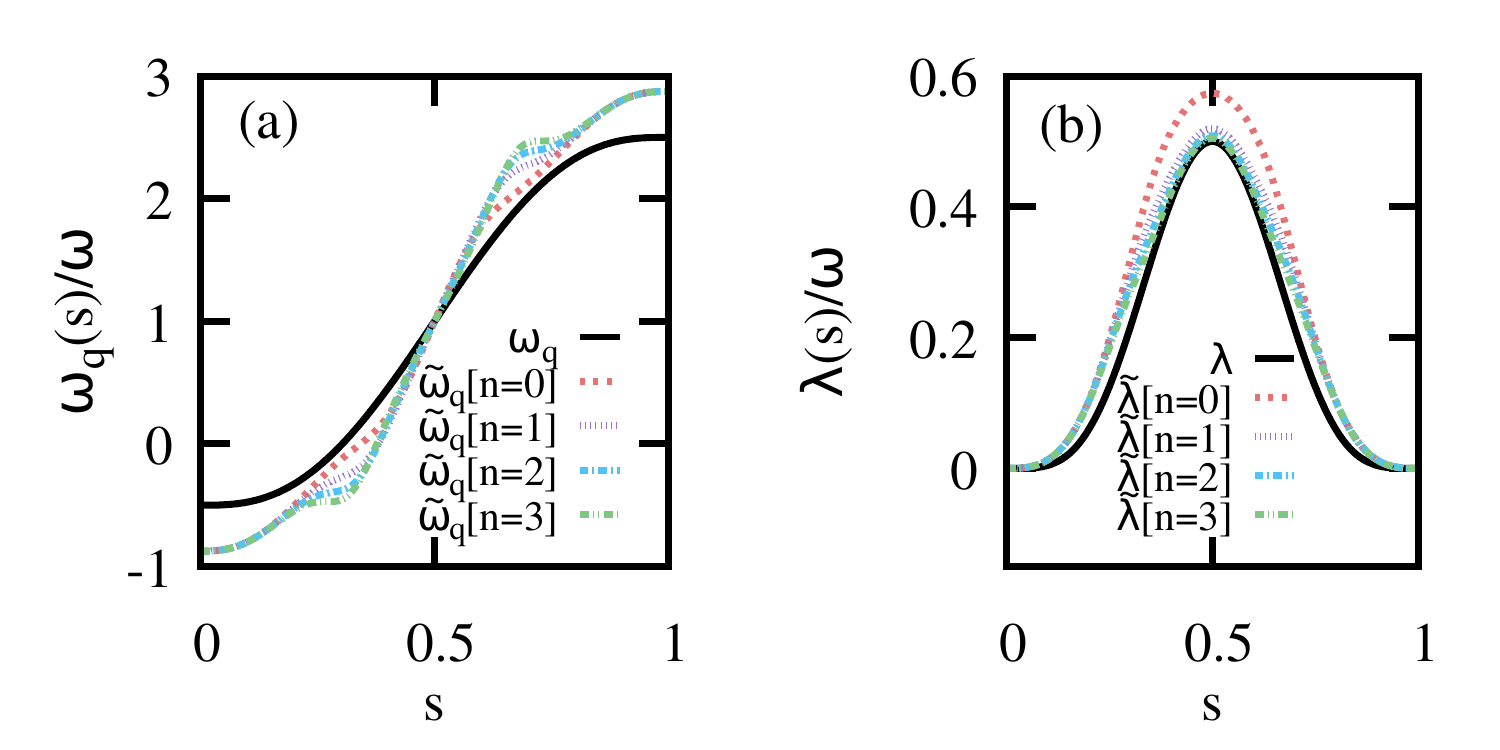}
\caption{\small{Panel (a) [(b)], solid black line: profile of the time-dependent parameter $\omega_q(s)$ [$\lambda(s)$] plotted against $s=t/\tau$ for $\omega\tau=10$, $\omega_q(0)=-\lambda_m=-\omega/2$ and $\omega_q(\tau)\!=\!5\omega/2$ with $\lambda_0=0$. STA is attained using $\tilde{\omega}_q(s)$ and $\tilde{\lambda}(s)$, shown here for the first four $n$-subspaces (dotted and dashed lines). }}
\label{fig1}
\end{figure}

\begin{figure*}[t]
\centering
\includegraphics[width=0.90\linewidth,angle=-0]{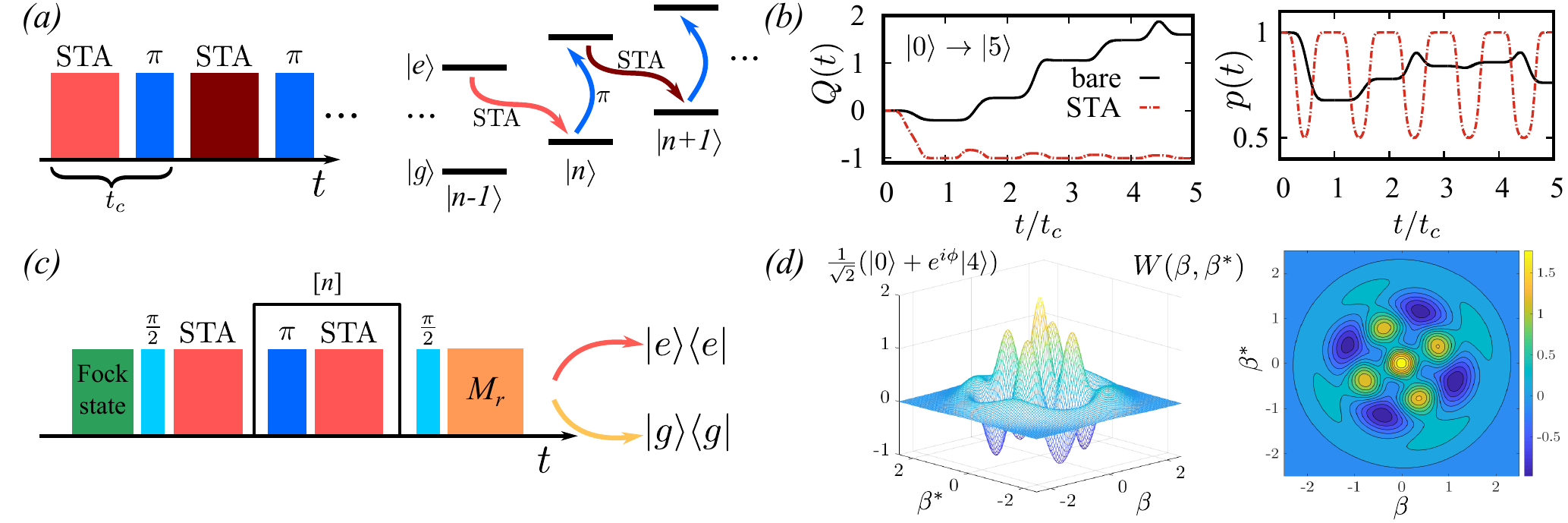}
\caption{\small{(a) Scheme for the generation of a Fock state $\ket{N}$ using STA and a $\pi$-pulse, and the transitions in the JC ladder. (b) Mandel parameter $Q(t)$ and purity $p(t)$ for the preparation of a $\ket{N=5}$ Fock state. We have used the STA drives $\tilde{\omega}_q(t)$ and $\tilde{\lambda}(t)$ with $\omega\tau=\omega t_\pi=5$ and $\omega\sigma_\pi=1$, with $\lambda_0=0$, $\lambda_m=\omega/4$ and $\omega_q(0)=3\omega_q(\tau)=3\omega/2$. 
  (c) Scheme for the preparation of a cat state based on Fock state preparation, $\pi/2$-pulses, STA and a projective measurement onto the spin. The pulses within the box can be performed $n$ times to generate superpositions comprising Fock states separated by $2(n+1)$. (d) Wigner function of the state $\rho_{\rm f}$  resulting from the application of the previous scheme to achieve $(\ket{0}+e^{i\phi}\ket{4})/\sqrt2$ with same parameters as above but $\omega \tau=30$.}}
\label{fig2}
\end{figure*}
%\frac{\tilde{\omega}_q(t)}{2}\sigma_z
We can however circumvent the difficulty in the implementation of an additional driving by performing a unitary transformation on $H_{\rm CD}^{\rm STA}(t)$ so as to obtain a local counterdiabatic Hamiltonian with the same form of the original $H_{\rm JC}(t)$~\cite{sup}, namely, $H_{\rm LCD}(t)\!=\!\tilde{\omega}_q(t)\sigma_z/2+\omega\adaga+\tilde{\lambda}(t)(a\sigma^++\adag\sigma^-)$, but with the new %frequency and coupling
parameters
\begin{equation}\label{eq:wqtilde}
  \tilde{\omega}_q(t)=\omega_q(t)-2\sqrt{(n+1)}\frac{\lambda(t)\dot{\theta}(t)-\theta(t)\dot{\lambda}(t)}{\theta^2(t)+\Omega_n^2(t)}, %\label{eq:ltilde}
 % \tilde{\lambda}(t)&=\left[\lambda^2(t)+\theta^2(t)%\frac{(\delta(t)\dot{\lambda}(t)-\lambda(t)\dot{\omega}_q(t))^2}{(\delta^2(t) + \Omega_n^2(t))^2} 
  %\right]^{1/2},
\end{equation}
and $ \tilde{\lambda}(t)=\sqrt{\lambda^2(t)+\theta^2(t)}$. %with $\theta(t)\!=\!\frac{\delta(t)\dot{\lambda}(t)-\lambda(t)\dot{\omega}_q(t)}{\Omega_n^2(t)+\delta^2(t)}$. 
Note that the driving must also fulfill $\ddot{\lambda}(0)\!=\!\ddot{\lambda}(\tau)\!=\!0$ and $\ddot{\omega}_q(0)\!=\!\ddot{\omega}_q(\tau)\!=\!0$ to ensure that $H_{\rm LCD}(t\!=\!0,\tau)\!=\!H_{\rm JC}(t\!=\!0,\tau)$. For that, we consider the protocols 
% Hamiltonians coincide with $H_{\rm JC}$
%\begin{equation}
%\begin{aligned}\label{eq:wqt}
 $\omega_q(t)\!=\!\omega_q(0)+10 \Delta\omega_q\, s^3 -15\Delta\omega_q s^4 +6\Delta\omega_qs^5$, %\label{eq:lt}
 and $\lambda(t)\!=\!(\lambda_m-\lambda_0)\cos^4\left[\pi \left(1+2 s\right)/2 \right]+\lambda_0$, 
%\end{aligned}
%\end{equation}
with $s\!=\!t/\tau$, $\Delta\omega_q\!=\!\omega_q(\tau)-\omega_q(0)$, and where $\lambda_0$ is the initial coupling constant, while $\lambda_m$ denotes its maximum value. As for a Landau-Zener problem, a  population transfer between $\ket{e,n}$ and $\ket{g,n+1}$ requires that $\omega_q(t)$ changes its sign during the evolution while $\lambda(t)\neq 0$ for some $t$ with $\lambda(0)\!=\!\lambda(\tau_q)\!=\!0$, which also applies to the modified frequencies.  In Fig.~\ref{fig1}, we illustrate the time-dependent behavior of the modified frequency and coupling parameters.
It is worth stressing that having control on $\omega_q(t)$ and $\lambda(t)$ allows for a perfect state transfer in the JC ladder for an arbitrary time $\tau$, while it is hindered when either $\dot{\omega}_q(t)\!=\!0$ or $\dot{\lambda}(t)\!=\!0$ $\forall t$. We remark that while $H_{\rm CD}^{\rm STA}(t)$ and $H_{\rm LCD}(t)$ perform in a similar manner, their associated energetic cost may differ~\cite{Abah:19}.

{\em Fock state generation.---}
As briefly mentioned, the implementation of the control $\tilde{\omega}_q(t)$ and $\tilde{\lambda}(t)$ allows for a perfect state transfer between $\ket{e,n}$ and $\ket{g,n+1}$, which can be used to generate an arbitrary Fock state $\ket{N}$ of the bosonic mode. Needless to say, for this specific case a time-independent evolution under $H_{\rm JC}$ may perform in a similar manner as our superadiabatic protocol~\cite{sup}, and the following example is given on a mere illustrative ground. We assume the initial state $\ket{e,0}$, then driven to $\ket{g,1}$ using a STA protocol. Upon a $\pi$-pulse on the spin, a STA is performed such that $\ket{g,2}$ is obtained. Concatenating this $N$ times, state $\ket{e,N}$ is achieved (cf. Fig.~\ref{fig2}(a)). In order to illustrate the performance of this protocol, we show the evolution of the Mandel parameter $Q(t)\!=\!(\langle n^2(t)\rangle-\langle n(t)\rangle^2)/\langle n(t)\rangle-1$ with $\langle n(t)\rangle = \bra{\psi(t)}\adaga\ket{\psi(t)}$, which accounts for the non-classicality of the resulting state. We also compute the purity $p(t)={\rm Tr}[\rho_s^2(t)]$ of the reduced two-level state $\rho_{\rm s}(t)={\rm Tr}_{b}[\rho(t)]$, with $\rho(t)=\ket{\psi(t)}\bra{\psi(t)}$ and ${\rm Tr}_b[\bullet]$ denoting the trace over the bosonic mode. Both $Q(t)$ and $p(t)$ showcase a perfect population transfer resulting from each STA+$\pi$-cycle of duration $t_c$ as we have $Q(t_c)=-1$ and $p(t_c)=1$ where $t_c=\tau+2t_\pi$ with  $\tau$ and $2t_\pi$ being the time spent in the STA evolution and  the $\pi$-pulse respectively.  The latter is modelled as a Gaussian function with standard deviation $\sigma_\pi$ [cf.~\cite{sup} for further details]. In Fig.~\ref{fig2}(b) we show the evolution of $Q(t)$ and $p(t)$ under STA for the target state $\ket{N=5}$ and compare them to the results obtained using the bare $H_{\rm JC}(t)$. The Mandel parameter $Q(t)$ clearly unveils the sub-Poissonian behavior of the boson statistics (i.e. $Q(t)<0$) for STA. Indeed, the STA protocol results in $Q(nt_c)=-p(n t_c)=-1$ and $p((n-1)t_c+\tau/2)=1/2$ for $n=1,2,\ldots 5$, since $\ket{\psi(n t_c)}=\ket{e,n}$ by construction~\cite{sup}. Under $H_{\rm JC}(t)$, the statistics becomes super-Poissonian, unless the protocol is performed sufficiently slow~\cite{sup}. %We make use of the frequencies $\tilde{\omega}_q(t)$ and $\tilde{\lambda}(t)$, which depend on $\omega_q(t)$ and $\lambda(t)$ with the values $\omega\tau=\omega t_\pi=5$ and $\omega\sigma_\pi=1$, with $\lambda_0=0$, $\lambda_m=\omega/4$ and $\omega_q(0)=3\omega_q(\tau)=3\omega/2$. 
%which depends on $\sigma_\pi$ and $t_\pi$

{\em Cat-state preparation.---} The so-called Schr\"odinger cat state, one of the paradigmatic examples of non-classical states, not only pose interest in fundamental quantum physics, but are highly valuable for quantum information processing applications. These states have been observed in numerous physical systems, including electronic~\cite{Clarke:88}, photonic~\cite{Auffeves:03,Ourjoumtsev:07,Leghtas:15}, and atomic degrees of freedom~\cite{Monroe:96,Wineland:13}. A scheme for the deterministic creation of Schr\"odinger's cat states has been recently demonstrated using a single three-level system trapped in an optical cavity~\cite{Hacker:19}. Nevertheless, it remains a big challenge to create superpositions of macroscopically distinct coherent states in nanomechanical systems~\cite{Xiang:13,Liao:16}. In order to realize a cat state we start from a particular Fock state (see discussion above) and first apply a $\pi/2$-pulse to split the quantum state in two different $n$-subspaces, upon which a fast state transfer (STA) is performed. In particular, for an initial state $\ket{e,N}$, the previous steps lead approximately to $\ket{e,N-1}+e^{i\phi}\ket{g,N+1}$ where $\phi$ denotes a relative phase acquired during the STA. Upon application of another $\pi/2$-pulse, followed by a projective measurement $M_r\!=\!\ket{r}\bra{r}\otimes \mathbb{I}_b$ onto the spin state ($r\in\{e,g\}$), the resulting bosonic state becomes $\ket{\psi_{N-1,N+1}}\sim(\ket{N-1}+e^{i\phi}\ket{N+1})/\sqrt2$. One can easily extend the previous sequence to generate cat states of a larger size by simply introducing $\pi$-pulses and additional STA evolution [cf. Fig.~\ref{fig2}(c)]. 
Note that, as the STA protocols depend on the addressed $n$-subspace, any given choice of $\tilde{\omega}_q(t)$ and $\tilde{\lambda}(t)$ cannot achieve perfect population transfer in two or more distinct $n$-subspaces simultaneously. However, this obstacle can be overcome by choosing parameters such that $\tilde{\omega}_q(t)$ and $\tilde{\lambda}(t)$ are similar in each of the required subspaces~\cite{sup}. As an example, in Fig.~\ref{fig2}(d) we show the Wigner function $W(\beta,\beta^*)=2{\rm Tr}[\rho_{\rm f} D(\beta)e^{i\pi\adaga}D^{\dagger}(\beta)]$~\cite{Davidovich:99}, with $D(\beta)=e^{\beta\adag-\beta^*a}$ the displacement operator, for an attained final state $\rho_{\rm f}$ involving $\ket{0}$ and $\ket{4}$, thus displaying the hallmarks of a cat state: distinguishable local-state components whose strong quantum interference results in negativity of $W(\beta,\beta^*)$. 
We benchmark the quality of our state-engineering protocol using the fidelity $F=\langle \psi_{\rm 0,4}|\rho_{\rm f}|\psi_{\rm 0,4} \rangle \gtrsim 0.999$ with $\ket{\psi_{\rm 0,4}}=\frac{1}{\sqrt{2}}(\ket{0}+e^{i\phi}\ket{4})$ and $\phi\approx \sqrt{2}\pi$. It is worth stressing that higher fidelities can be achieved depending on the choice of the parameters, while a time-independent evolution leads to $F\approx 0.7$~\cite{sup}.

\begin{figure}[!]
\centering
\includegraphics[width=0.8\linewidth]{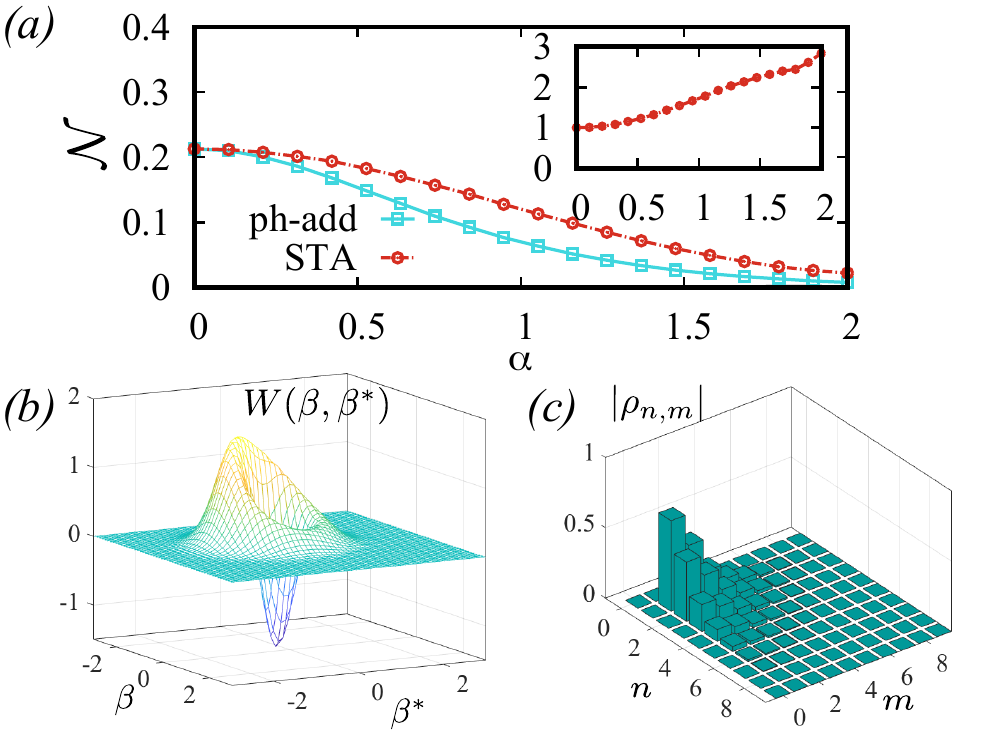}
\caption{\small{(a) Non-classicality $\mathcal{N}$ after a STA evolution removing the vacuum of $\ket{e,\alpha}$ and the value associated to photon addition $\ket{\psi_{\rm ph-add}}$. The inset shows the ratio $\mathcal{N}_{\rm STA}/\mathcal{N}_{\rm ph-add}$. (b) Wigner function of the mode state after the protocol to remove the vacuum for $\alpha=3/4$, and (c) its associated state for the first $10$ Fock states,  $|\rho_{n,m}|=|\langle n |\rho_{\rm f}|m\rangle|$ (same parameters as in Fig.~\ref{fig2}(b) with $\omega\tau=8$).}}
\label{fig3}
\end{figure}

{\em Photon-shifted states.---} 
An interesting class of non-classical states is generated by the combination of addition and subtraction of bosonic excitations~\cite{Kim:08}. Their most basic embodiments consists of the addition or subtraction of a single quantum, which results in $\ket{\psi_{\rm ph-add}}\propto \adag \ket{\psi}$ and $\ket{\psi_{\rm ph-sub}}\propto a \ket{\psi}$, respectively. These arithmetic operations are important in quantum-based technologies~\cite{Browne:17,Lee:95,Wenger:04,Kim:05,Parigi:07,Zavatta:09,Barnett:18}. %In optics, these techniques require the clever use of beam splitters, non-linear media and post-selection to generate such non-classical states.  Note that while photon addition always produces a non-classical field state, photon subtraction yields a final non-classical field only if the original state was already non-classical \cite{Lee:95,Wenger:04,Kim:05,Parigi:07,Zavatta:09,Barnett:18}.  
Building on our scheme, we now show how to produce non-classical states --  which we term photon-shifted states -- achieved by transferring the population of the field vacuum to excitation-bearing Fock states. Such state manipulation has recently been demonstrated in a trapped ion system via an anti-JC interaction~\cite{Um:16}. The step forward embodied by our proposal is that photon-shifted states can be generated in a fast and controllable manner using a single STA driving as follows. Let us consider an initial coherent state $\ket{e,\alpha}=D(\alpha)\ket{e,0}$. By applying a STA driving for the $n=0$ subspace, one removes exactly the population of the vacuum state and shifts it to $\ket{1}$. The scheme is repeated, after a $\pi$-pulse, to progressively transfer population to higher-excitation Fock states. Remarkably, provided $|\alpha| \lesssim 1$, the previous protocol approximately corresponds to a photon addition. However, our scheme yields in general states that are more non-classical than $\ket{\psi_{\rm ph-add}}$. To prove such claim, we use the negativity of the Wigner function $\mathcal{N}=\frac{1}{2\pi}\int d^2\beta (|W(\beta,\beta^*)|-W(\beta,\beta^*))$~\cite{Kenfack:04}, which is shown in Fig.~\ref{fig3}(a) against the value of $\alpha$ of the initial state. Clearly, a photon-shifted state achieves a larger value of ${\cal N}$ -- and thus more non-classicality -- than a photon-added state.  The removal of the vacuum and population-shift has a profound impact on the Wigner function of the mode and on the corresponding state $\rho_{\rm f}$ (cf. Fig.~\ref{fig3}(b)-(c)).
Although not explicitly shown, similar results are obtained for other initial field states, such as thermal states~\cite{sup}.

{\em Robustness.---} As our scheme is built on STA protocols allowing for short evolution times, the method is naturally robust against decoherence effects. In particular, we can achieve a desired non-classical target state under a broad range of noise rates of typical decoherence processes, such as spin dephasing, spontaneous emission, mode heating and damping (see~\cite{sup} for more details and numerical results). Moreover, we have checked the robustness of the method to pulse-shape variations, which is a relevant step towards the actual implementation of the STA protocols. In order to evaluate the effect of such imperfections, we considered the preparation of $\ket{\psi_{0,4}}$ for $\tilde{\omega}_q(t)$ and $\tilde{\lambda}_q(t)$ approximated as $\tilde{x}_{\rm F}(t)\!=\!\sum_{k=0}^{N_{\rm F}}c_k\cos(k\omega_{\rm F} t)+s_k\sin(k\omega_{\rm F} t)$ with $\tilde{x}\in\{\tilde{\omega}_q,\tilde{\lambda}\}$. The cat state shown in Fig.~\ref{fig2}(d) is achieved with $F\gtrsim 0.99$ already for $N_{\rm F}=2$ (see~\cite{sup} for further examples and details). This demonstrates the robustness of the proposed protocols.

{\em Experimental feasibility.---}
Our scheme can be realized in a variety of physical systems where a JC interaction between a two-level system and a bosonic field can be controlled, such as in superconducting qubits~\cite{Forn:17,Kockum:19}, trapped ions~\cite{Leibfried:03,Haffner:08,Pedernales:15} or spin-mechanical systems~\cite{Rae:04,Treutlein:14,Abdi:17}, among others. Here we focus on an ion-trap implementation~\cite{Um:16,Lv:17,Lv:18}. In this setup, a well-controllable qubit can be encoded on the two magnetically-insensitive hyperfine states of the $S_{1/2}$ manifold of a  ${}^{171}{\rm Yb}^+$ ion~\cite{Olmschenk:07}, whose frequency is $\omega_{\rm hf}/2\pi\!=\!12.6428$ GHz. The trapped-ion is confined in a harmonic potential with frequency $\omega_{X}/2\pi\approx2$ MHz~\cite{Um:16,Lv:17,Lv:18}, such that the free Hamiltonian reads $H_0\!=\!\omega_{\rm hf}\sigma_z/2+\omega_X\adaga$. Applying two counter-propagating Raman laser beams, the internal levels of the ion can be coupled with the vibrational mode as $H_{\rm int}\!=\!\Omega \cos(\Delta k x-\omega_l t-\phi)\sigma_x$, where $\Omega$, $\Delta k$, $\omega_l$ and $\phi$ are the Rabi frequency, net wave vector on the $x$-axis, frequency and phase of the laser fields, respectively, while $x\!=\!(2m\omega_X)^{-1/2}(a+\adag)$ is the position operator of the ion with mass $m$. In an interaction picture with respect to $H_0$, upon the  optical and vibrational rotating wave approximations, and within the Lamb-Dicke regime, one obtains $H_{\rm int}^I\!=\!U_0^{\dagger}(t)H_{\rm int}U_0(t)\approx \lambda (a\sigma^+e^{i\delta t}+{\rm H.c.})$ with $U_0(t)\!=\!e^{-itH_0}$, $\lambda=\Omega\Delta k (2m\omega_X)^{-1/2}/2$ and $\phi\!=\!\pi/2$, and where we have selected $\omega_l=\omega_{\rm hf}-\omega_X-\delta$ with $\delta\ll \omega_X$~\cite{Um:16,Lv:17,Lv:18}. Note that $H_{\rm int}^I$ already corresponds to $H_{\rm JC}(t)$ (cf. Eq.~\eqref{eq:HJC}) in the interaction picture of $\omega_q(t)\sigma_z/2+\omega\adaga$, requiring a detuning $\delta(t)=\omega_q(t)-\omega$ and a modulated laser intensity $\Omega(t)$, such that the proposed protocols $\lambda(t)$ and $\omega_q(t)$ can be realized. Indeed, $\lambda/2\pi\approx12.5$ kHz and $\delta(t)$ can be varied within $|\delta(t)|/(2\pi)\sim 0-100$ kHz while ensuring the correct functioning of the required approximations~\cite{Pedernales:15,Um:16,Lv:17,Lv:18}. A possible set of realistic parameters to implement the scheme is given by $\omega/2\pi\approx50$ kHz, such that $\lambda_m\approx\omega/4$, which leads to $\tau\approx 0.2$ ms for $\omega\tau\!=\!10$. For such short $\tau$, decoherence effects are not expected to play a relevant role~\cite{Um:16,Lv:18}, and one may still rely on suitable dynamical decoupling schemes to further protect the system against decoherence processes~\cite{Lidar:14,Souza:12,Cai:12,Puebla:16njp,Puebla:17pra,Puebla:18jmod}.

{\em Conclusions.---} We have developed a general framework for a fast, robust and accurate preparation of non-classical states in spin-boson systems that are highly desirable in, for example, quantum information processing tasks~\cite{Kim:08} and fundamental physics inquiries~\cite{Hornberger:12,Bassi:13}. In particular, the proposed pulses allow for a perfect state transfer in a JC model, which are built relying on STA. As an illustration of the potential and versatility of the method, we show how to generate arbitrary Fock states and cat states. In addition, we show how to obtain a class of photon-shifted states where  the vacuum population can be removed, thus similar to photon addition but featuring more non-classicality. These protocols, intrinsically robust against decoherence thanks to their arbitrarily short evolution time, are also resilient to imperfect implementation or modifications to their actual shape profiles. Our results may open new routes and possibilities for an efficient preparation of non-classical states in a variety of settings, amenable for their experimental realization in state-of-the-art setups.

\begin{acknowledgments} The authors are grateful to K. Kim for useful discussions and acknowledge the support by the SFI-DfE Investigator Programme (grant 15/IA/2864), the Royal Commission for the Exhibition of 1851, the H2020 Collaborative Project TEQ (Grant Agreement 766900), the Leverhulme Trust Research Project Grant UltraQuTe (grant nr. RGP-2018-266) and the Royal Society Wolfson Fellowship (RSWF/R3/183013), and the Royal Society International Exchanges scheme (IEC/R2/192220)
\end{acknowledgments}

%\bibliographystyle{mystyle.bst}
%\bibliographystyle{apsrev4-1.bst}
%\bibliography{qrm_heat_engine.bib,references.bib,hierarchy.bib,notes.bib}
%\bibliography{QSE.bib}


%\end{document}

\widetext
\clearpage
\begin{center}
\textbf{\large Supplemental Material\\ Quantum state engineering by shortcuts-to-adiabaticity in interacting spin-boson systems}
\end{center}
%%%%%%%%%% Merge with supplemental materials %%%%%%%%%%
%%%%%%%%%% Prefix a "S" to all equations, figures, tables and reset the counter %%%%%%%%%%
\setcounter{equation}{0}
\setcounter{figure}{0}
\setcounter{table}{0}
\setcounter{page}{1}
\makeatletter
\renewcommand{\theequation}{S\arabic{equation}}
\renewcommand{\thefigure}{S\arabic{figure}}
\renewcommand{\bibnumfmt}[1]{[S#1]}
\renewcommand{\citenumfont}[1]{S#1}

\begin{center}
  Obinna Abah${}^*$, Ricardo Puebla${}^\dagger$ and Mauro Paternostro\\\vspace{0.25cm}{\em \small{
      Centre for Theoretical Atomic, Molecular, and Optical Physics,\\
      School of Mathematics and Physics, Queen's University, Belfast BT7 1NN, United Kingdom}}\\
  \small{(Dated: \today)}
\end{center}

\section{STA for time-dependent two-level system}\label{sec:S1}
We consider a generalized time-dependent system Hamiltonian of the form
\begin{align}
H_0(t) = \frac{\Delta(t)}{2}\sigma_x+\frac{\lambda(t)}{2}\sigma_z, 
\label{Seq1}
\end{align}
where the time-independent of $\Delta(t) \equiv \Delta$ ($\lambda(t) \equiv \lambda$) corresponds to the Landau-Zener (Rosen-Zener) model which is well studied in many physical settings. %\cite{Rosen:32}%This kind of Hamiltonian is actually what Chen et al. considered in \cite{Chen2010PRLb}.
From Berry's formulation, the resulting counterdiabatic Hamiltonian reads as~\cite{SMBerry2009JPA}
\begin{align}\label{eq:SCD}
H_{\text{CD}}(t) = i\sum_n\left[\ket{\partial_t n(t)}\bra{n(t)} - \bra{n(t)}\partial_t \ket{n(t)} \ket{n(t)}\bra{n(t)}\right],
\end{align}
where $\ket{n(t)}$ is the eigenstate of the original Hamiltonian $H_0(t)$ and the $H_{\rm CD}(t)$ is Hermitian and non-diagonal in the $\ket{n(t)}$ basis. Note that the Eq.~\eqref{eq:SCD} is equivalent to the expression given in the main text, $H_{\rm CD}(t)=i\sum_n [\partial_t\Phi_n(t),\Phi_n(t)]$ with $\Phi_n(t)=\ket{n(t)}\bra{n(t)}$ the projector onto the instantaneous $n$-eigenstate $\ket{n(t)}$ of $H_0(t)$. For the Hamiltonian above, Eq.~\eqref{Seq1} the counterdiabatic Hamiltonian has the form
\begin{align}
H_{\rm CD}(t) = \frac{\lambda(t)\dot{\Delta}(t) - \Delta(t)\dot{\lambda}(t)}{2 \left(\lambda^2(t) + \Delta^2(t) \right)} \sigma_y \equiv \frac{\theta_a(t)}{2}.
\label{eq:SqCD}
\end{align}
The total Hamiltonian $H(t) = H_0(t) + H_{\rm CD}(t)$ that ensures perfect transfer at any given time becomes
\begin{align}
H(t) = \frac{\Delta(t)}{2}\sigma_x+\frac{\lambda(t)}{2}\sigma_z +  \frac{\lambda(t)\dot{\Delta}(t) - \Delta(t)\dot{\lambda}(t)}{2 \left(\lambda^2(t) + \Delta^2(t) \right)} \sigma_y.
\end{align}
To circumvent the difficulty in the implementation of additional $\sigma_y$-field driving, one make a time-dependent unitary transformation to the total Hamiltonian $H(t)$ such that
\begin{align}
|\tilde{\psi}(t)\rangle = U^\dagger(t)\ket{\psi(t)},
\end{align}
 choosing $U(t) = e^{-i f(t)\sigma_z}$. The modified Hamiltonian is obtained using
 \begin{align}
 H_{\mathrm{LCD}}(t) = U^\dagger(t)(H(t) - i \hbar \dot{U}(t) U^\dagger(t)) U(t),
 \end{align}
 which can be written as
 \begin{align}
 H_{\mathrm{LCD}}(t) = \left(\frac{\Delta(t)}{2} \cos(2 f(t)) + \frac{\theta_a(t)}{2} \sin(2f(t))\right) \sigma_x + \left( \frac{\theta_a(t)}{2} \cos(2 f(t)) - \frac{\Delta}{2}\sin(2 f(t)) \right) \sigma_y + \left(\frac{\lambda(t)}{2} - \dot{f}(t)\right)\sigma_z.
 \end{align}
 Hence, the $\sigma_y$-term is eliminated when $f(t)$ is of the form
 \begin{align}
 f(t) = \frac{1}{2} \arctan\left(\frac{\theta_a(t)}{\Delta(t)}\right). 
 \end{align}
  Using the above expression, the resulting LCD Hamiltonian reads as
 \begin{align}
 H_{\rm LCD}(t) = \frac{\Delta(t)}{2} \sqrt{1 + \frac{\theta_a^2(t)}{\Delta^2(t)}} \sigma_x + \left(\frac{\lambda(t)}{2} - \frac{\Delta(t) \dot{\theta}_a(t) - \theta_a(t)\dot{\Delta}(t)}{2(\theta_a^2(t) +\Delta^2(t))}\right) \sigma_z.
 \label{eq:SqLCD}
 \end{align}
 %We will now use these equations (CD and LCD) to write the general form of JC model.

\section{STA for the Jaynes-Cummings model}
Let now consider a time dependent Jaynes-Cummings  model of the form ($\hbar =1$)
\begin{align}
H_{\rm JC}(t)=\frac{\omega_q(t)}{2}\sigma_z+\omega\adaga+\lambda(t)(a\sigma^++\adag \sigma^-).
\end{align}
%As we have seen, $H_{\rm JC}$ decouples in blocks with fixed number of excitations $n$, in the dressed basis $\ket{e,n}$ and $\ket{g,n+1}$,
%\begin{align}
%H_{n}=\frac{(2n+1)\omega}{2}\mathbb{I}+\frac{\delta(t)}{2}\bar{\sigma}_z+g(t)\sqrt{n+1}\bar{\sigma}_x
 % \end{align} 
Since the total number of excitations $N_e\equiv \ket{e}\bra{e}+\adaga$ is conserved, we can diagonalize the Hamiltonian in blocks, in the subspace $S_n$ spanned by $\left\{\ket{e,n},\ket{g,n+1} \right\}$ with $n=0,1,\ldots$. Then,
\begin{align}
H_{n}(t)=\left( \begin{matrix} \bra{e,n}H_{\rm JC}(t)\ket{e,n} &\bra{e,n}H_{\rm JC}(t)\ket{g,n+1} \\\bra{g,n+1}H_{\rm JC}(t)\ket{e,n} & \bra{g,n+1}H_{\rm JC}(t)\ket{g,n+1} \end{matrix} \right)= \frac{1}{2} \left(\begin{matrix}  \delta(t)+(2n+1)\omega & 2\lambda(t)\sqrt{n+1} \\2\lambda(t)\sqrt{n+1} & -\delta(t)+(2n+1)\omega \end{matrix}\right),
  \end{align}
such that $H_{\rm JC}(t)=-\omega_q(t)/2 \ket{g,0}\bra{g,0}+\bigoplus_nH_n(t)$ and with $\delta(t)=\omega_q(t)-\omega$. The previous Hamiltonian can be mapped to that of a generalized time-dependent two-level system, simply by defining $\bar{\sigma}^-=\ket{g,n+1}\bra{e,n}$, $\bar{\sigma}^+=\ket{e,n}\bra{g,n+1}$, $\bar{\sigma}_z=\ket{e,n}\bra{e,n}-\ket{g,n+1}\bra{g,n+1}$, such that
\begin{align}
H_{n}(t)=\frac{(2n+1)\omega}{2}\mathbb{I}+\frac{\delta(t)}{2}\bar{\sigma}_z+\lambda(t)\sqrt{n+1}\bar{\sigma}_x.
  \end{align}
 Moreover, rotating  $\pi/2$ about the $y$-axis, $\bar{\sigma}_z\rightarrow \bar{\sigma}_x$ and $\bar{\sigma}_x\rightarrow -\bar{\sigma}_z$, we obtain
\begin{align}\label{eq:HnLZ}
  H_{n}(t)=\frac{(2n+1)\omega}{2}\mathbb{I}+\frac{\delta(t)}{2}\bar{\sigma}_x - \frac{2 \lambda(t) \sqrt{n+1}}{2}\bar{\sigma}_z.
  \end{align}
 For $\lambda(t) =0$, the eigenstates are $\ket{\phi_{1,2}(\lambda=0)}=\ket{\pm}$ in the basis of $\bar{\sigma}_x$. From Eq.~(\ref{eq:HnLZ}), we can already notice that Eqs.~(\ref{eq:SqCD}) and~(\ref{eq:SqLCD}) are valid upon the identification of $\Delta\rightarrow \delta(t)$ and $\lambda(t)\rightarrow -2 \lambda(t)\sqrt{n+1}$. 

Following the definition of control term to suppress nonadiabatic transitions in~\cite{SMDemirplak2003,SMBerry2009JPA}
\begin{equation}
\label{JCcd}
\begin{aligned} 
H_\text{CD}(t) &= i \sum_{n,\sigma=\pm}  (\ket{\partial_t (n,\sigma(t))} \bra{n,\sigma (t)}-  \langle n,\sigma (t)|\partial_t (n,\sigma(t))\rangle \ket{n,\sigma (t)}\bra{n,\sigma (t)}), 
\end{aligned}
\end{equation}
with $\ket{n,\sigma (t)}$ denoting the dressed-atom eigenstates of the $H_{\rm JC}(t)$. The resulting  counterdiabatic Hamiltonian reads as
\begin{equation}
H_\text{CD}(t)= \frac{\dot{\lambda}(t)\delta(t) - \lambda(t)\dot{\omega}_q(t)}{\delta^2(t)+ \Omega_n^2(t)}(i\adag\sigma^--ia\sigma^+).
\end{equation}
This additional driving in the $\bar{\sigma}_y$ direction suppresses non-adiabatic excitations allowing for an arbitrarily fast adiabatic evolution.

For the LCD, we proceed in a straightforward manner as illustrated in  Section S1,  the LCD Hamiltonian is given by
\begin{align}
H_{\rm LCD}(t)=\frac{\tilde{\omega}_q(t)}{2}\sigma_z+\omega\adaga+\tilde{\lambda}(t)(a\sigma^++\adag\sigma^-),
  \end{align}
but with modified qubit frequency and coupling parameter, 
\begin{align}\label{eqS:wqtilde}
  \tilde{\omega}_q(t)&=\omega_q(t)-2\sqrt{(n+1)}\frac{\lambda(t)\dot{\theta}(t)-\theta(t)\dot{\lambda}(t)}{\theta^2(t)+\Omega_n^2(t)},\\ \label{eqS:ltilde}
  \tilde{\lambda}(t)&=\left[\lambda^2(t)+\frac{(\delta(t)\dot{\lambda}(t)-\lambda(t)\dot{\omega}_q(t))^2}{(\delta^2(t) + \Omega_n^2(t))^2} \right]^{1/2},
\end{align}
with $\theta(t)=\frac{\delta(t)\dot{\lambda}(t)-\lambda(t)\dot{\omega}_q(t)}{\Omega_n^2(t)+\delta^2(t)}$. 
For a perfect state transfer, the driving protocols must also fulfill $\ddot{\lambda}(0)\!=\!\ddot{\lambda}(\tau)\!=\!0$ as well as $\ddot{\omega}_q(0)\!=\!\ddot{\omega}_q(\tau)\!=\!0$. For that, we consider the following protocols
% Hamiltonians coincide with $H_{\rm JC}$
\begin{align}\label{eqS:wqt}
  \omega_q(t)&=\omega_q(0)+10 \Delta\omega_q\, s^3 -15\Delta\omega_q s^4 +6\Delta\omega_qs^5,\\ \label{eqS:lt}
  \lambda(t)&=(\lambda_m-\lambda_0)\cos^4\left[\pi \left(1+2 s\right)/2 \right]+\lambda_0
\end{align}
with $s\!=\!t/\tau$, $\Delta\omega_q\!=\!\omega_q(\tau)-\omega_q(0)$, and $\lambda_0$ is the initial coupling constant, while $\lambda_m$ denotes its maximum value.

% In Fig.~\ref{fig1}, we illustrate the time-dependent behavior of the modified frequency and coupling parameters. It is worth stressing that having a time-dependent controls on $\omega_q(t)$ a $\lambda(t)$ allows for a perfect state transfer in the JC ladder for an arbitrary time $\tau$, while it is hindered when either $\dot{\omega}_q(t)=0$ or $\dot{\lambda}(t)=0$ $\forall t$ (see~\cite{sup} for further details).
% \begin{figure}
%\centering
%\includegraphics[width=0.66\linewidth,angle=-0]{fig1}
%\caption{\small{Profile of the time-dependent parameters, $\omega_q(s)$ and $\lambda(s)$ in (a) and (b), respectively, plotted with solid (black) lines and $s=t/\tau$ for $\omega\tau=10$, $\omega_q(0)=-\lambda_m=-\omega/2$ and $\omega_q(\tau)=5\omega/2$ with $\lambda_0=0$. STA is attained using $\tilde{\omega}_q(s)$ and $\tilde{\lambda}(s)$, Eqs.~\eqref{eq:wqtilde} and~\eqref{eq:ltilde}, shown here for the first four $n$-subspaces (dotted and dashed lines). }}
%\label{fig1}
%\end{figure}

%\RP{On performance of the STA for $\lambda_0\neq 0$}

\section{$\pi$ and $\pi/2$-pulses}\label{app:pi}
The shape of the pulse is considered here to be Gaussian. The state during the application of the pulse evolves under the Hamiltonian 
\begin{align}\label{eqa:Hpi}
H_{\pi}(t)=\frac{1}{2}\sqrt{\frac{\pi}{2}}e^{-\frac{(t-t_\pi)^2}{2\sigma_\pi^2}}\sigma_x,
\end{align}
and thus it depends on the time $t_\pi$ and the standard deviation $\sigma_\pi$. One can clearly see that the evolution under $H_{\pi}(t)$ produces a $\pi$-pulse for a total time $(T-t_\pi)/\sigma_\pi\gg 1$. For a $\pi/2$-pulse, the amplitude in Eq.~\eqref{eqa:Hpi} is simply reduced by a factor $2$.

\begin{figure}
\centering
\includegraphics[width=0.66\linewidth,angle=-0]{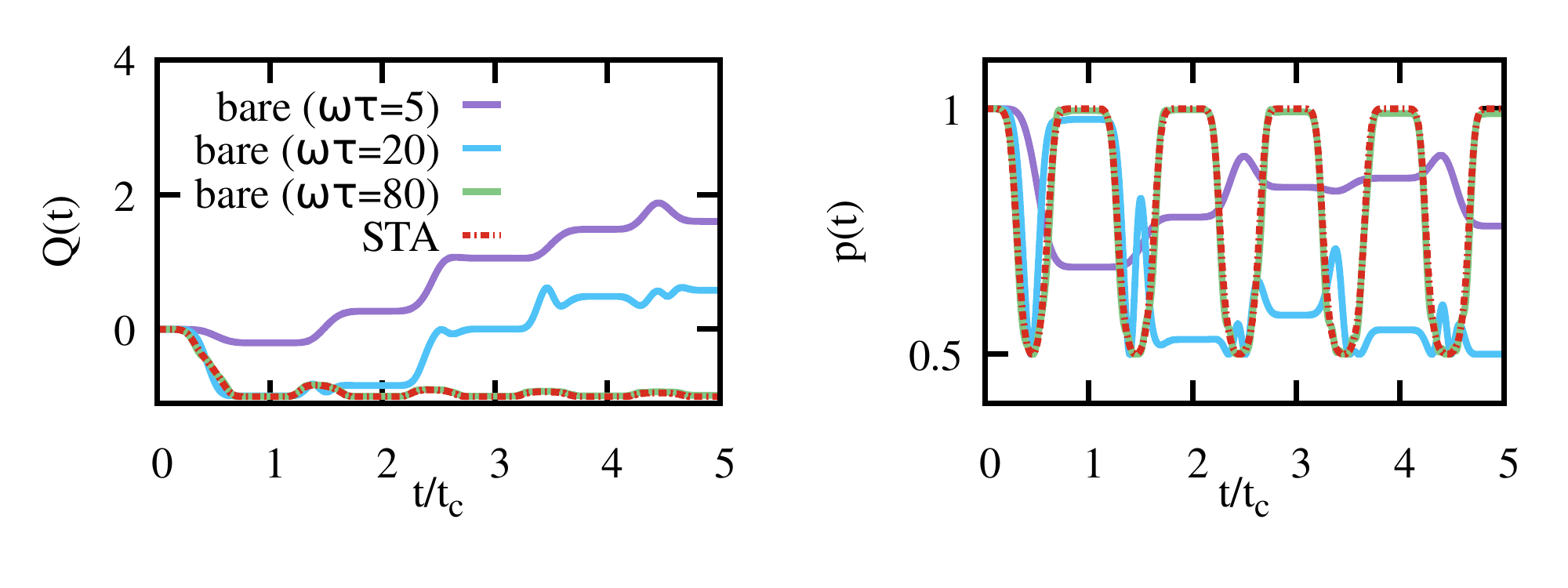}
\caption{\small{Evolution of the Mandel parameter $Q(t)$ (left) and purity of the reduced spin state $p(t)={\rm Tr}[\rho_s^2(t)]$ (right) as a function of the protocol time, where $t_c=\tau+2t_\pi$. Here we choose the same parameters as considered in the Fig. 2(b) of the main text,  $\omega\sigma_\pi=1$, with $\lambda_0=0$, $\lambda_m=\omega/4$ and $\omega_q(0)=3\omega_q(\tau)=3\omega/2$, while $\omega\tau=\omega t_\pi$ is denoted in the legend ($5$ (violet), $20$ (blue) and $80$ (green) for bare evolutions). The STA (dashed red) is independent of the chosen $\tau$. Note that for very long protocol times, $\omega\tau\rightarrow \infty$, the bare evolution leads to the one obtained for the STA. For the considered parameters here, we obtain a good overlap between the STA and the bare evolution for $\omega\tau=80$, although $Q(5t_c)\approx -0.98$ and $p(5t_c)\approx 0.99$ for the latter.}}
\label{figBareevol}
\end{figure}

\section{Comparison between bare and STA protocols}
The bare evolution  leads to an adiabatic result in the long time limit, i.e. provided $\omega\tau\rightarrow\infty$. In this manner, the results for the STA shown in Fig. 2(b) in the main text, will be eventually achieved using a bare evolution with $\omega\tau\rightarrow\infty$. This is exemplified in Fig.~\ref{figBareevol}, which is similar to Fig. 2(b) in the main text but including three different examples of bare evolutions, namely, $\omega\tau=5$, $20$ and $80$, aiming to prepare the Fock state $\ket{N=5}$ (see Fig.~\ref{figBareevol} for the used parameters). By definition, the STA provides the exact adiabatic result, regardless of $\omega\tau$.

Recall that, in order to prepare a Fock state $\ket{N}$, we perform the transfer from $\ket{\psi(0)}=\ket{e,n}$ to $\ket{\psi(\tau)}=\ket{g,n+1}$ (i.e. an adiabatic evolution in a time $\tau$). Hence, at half way the state reads $\ket{\psi(\tau/2)}=\frac{1}{\sqrt{2}}(\ket{e,n}\pm\ket{g,n+1})$. Hence, the purity of the reduced spin state is readily given by $p(\tau/2)={\rm Tr}[\rho_s^2(\tau/2)]=1/2$. This quantity indicates that the STA evolution sweeps across the $\ket{e,n} \leftrightarrow \ket{g,n+1}$ transition, as it can be seen in Fig. 2(b) and here in Fig.~\ref{figBareevol}.

\section{Impact of thermal boson states into Fock state preparation}
\begin{figure}
\centering
\includegraphics[width=1\linewidth,angle=-0]{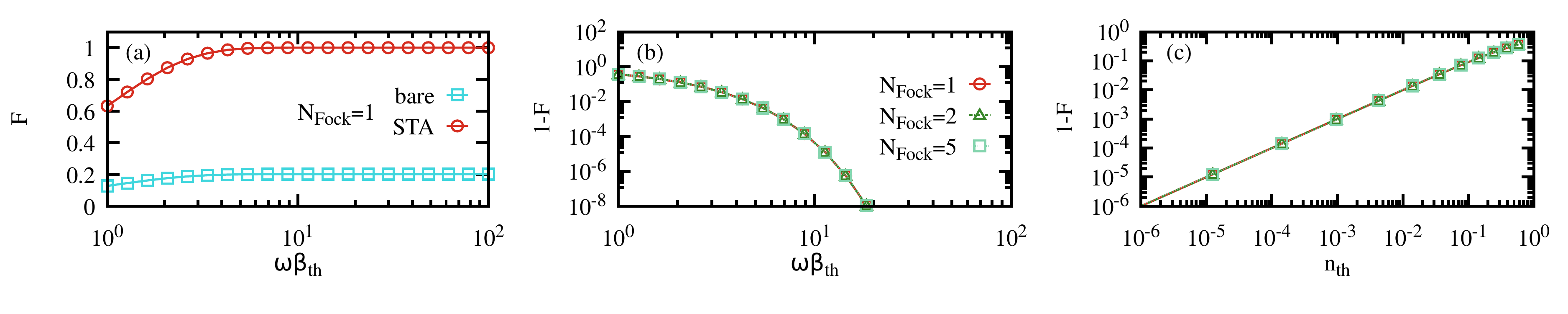}
\caption{\small{(a) Fidelity between the final state and the sought Fock state, $\ket{1}$, after an evolution using the bare (blue squares) and STA (red circles) protocols starting with a thermal state, $\rho(0)=\ket{e}\bra{e}\otimes\rho_{\rm th}$, with temperature $\beta_{\rm th}^{-1}$. For large temperatures, small $\beta_{\rm th}$, the fidelity $F$ is reduced. (b) Plot of the infidelity $1-F$ to better illustrate how $F$ is deteriorated for small $\beta_{\rm th}$ values. Note that the results do not depend on the desired Fock state, shown here for $\ket{1}$, $\ket{2}$ and $\ket{5}$. (c) Same results as in (b) but as a function of the thermal boson number of the initial state, $n_{\rm th}=(e^{\omega\beta_{\rm th}}-1)^{-1}$.  The parameters used for the simulation are $\lambda_0=0$, $\lambda_m=\omega/4$, $\omega_q(0)=3\omega_q(\tau)=3\omega/2$ and $\omega\tau=5$.}}
\label{figThermalFock}
\end{figure}

The method explained in the main text for Fock state preparation relies on an initial vacuum state of the field, namely, $\ket{\psi(0)}=\ket{e,0}$. Here we analyze how the resulting state upon a STA evolution deviates from the aimed Fock state when the initial field state finds itself in a thermal state, $\rho(0)=\ket{e}\bra{e}\otimes \rho_{\rm th}$ with $\rho_{\rm th}=\frac{e^{\beta_{\rm th}\omega\adaga}}{Z}\ket{n}\bra{n}$ with $Z={\rm Tr}[e^{-\beta_{\rm th}\omega\adaga}]$ and where $\beta_{\rm th}$ stands for the inverse of the temperature. Recall that the number of boson in a thermal state is given by $n_{\rm th}=(e^{\omega\beta_{\rm th}}-1)^{-1}$. We make use of the protocol explained in the main text for the preparation of an arbitrary Fock state. 
In Fig.~\ref{figThermalFock} we show the results for the fidelity $F$ between the final state and the desired Fock state $N_{\rm Fock}$, namely, $F=\bra{N_{\rm Fock}}\rho_{\rm f}\ket{N_{\rm Fock}}$ where $\rho_{\rm f}$ is the reduced field state upon the protocol is completed (of total duration $N_{\rm Fock}\tau$). We choose $\lambda_0=0$, $\lambda_m=\omega/4$, $\omega_q(0)=3\omega_q(\tau)=3\omega/2$ and $\omega\tau=5$ for a single STA/bare evolution. In Fig.~\ref{figThermalFock}(a) we compare the obtained $F$ to prepare a $\ket{N_{\rm Fock}=1}$ state using the bare Hamiltonian and the STA protocol as a function of the temperature. For large $\beta_{\rm th}$ values the initial state is essentially a vacuum state so that $F\rightarrow 1$ for $\beta_{\rm th}\rightarrow \infty$. As the STA protocol brings exactly the population from $\ket{e,n}$ into $\ket{g,n+1}$, the deterioration in the fidelity to bring $\rho_{\rm th}$ into $\ket{N_{\rm Fock}=1}$ is equal to when higher Fock states are sought. This is illustrated in Fig.~\ref{figThermalFock}(b) for $\ket{1}$, $\ket{2}$ and $\ket{5}$.  Finally, for small $1-F$ and $n_{\rm th}$ values,  we observe that $1-F$ scales linearly with the number of bosons in the initial thermal state $n_{\rm th}$, namely, $1-F\sim n_{\rm th}$  (cf. Fig.~\ref{figThermalFock}(c)).  

\begin{figure}
\centering
\includegraphics[width=1\linewidth,angle=-0]{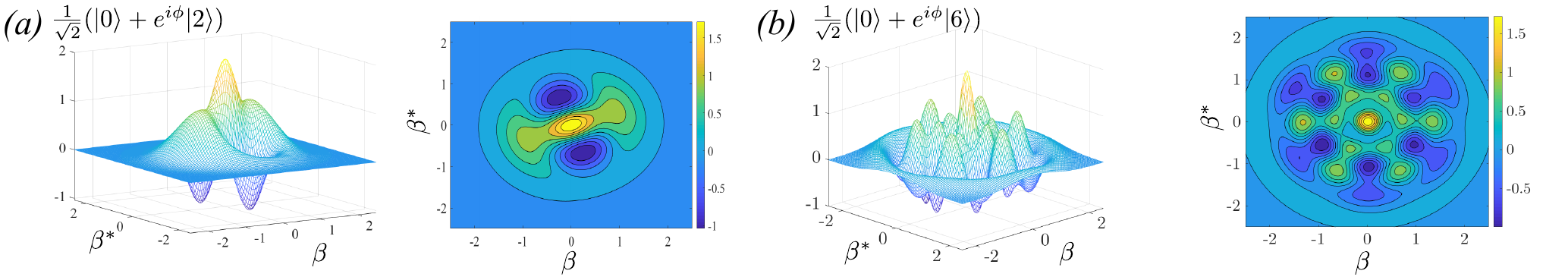}
\caption{\small{Wigner functions $W(\beta,\beta^*)$ obtained for two different cases using the STA protocols, (a) $\ket{\psi_{0,2}}$ and (b) $\ket{\psi_{0,6}}$. Again, $\lambda_0=0$, $\lambda_m=\omega/4$, $\omega_q(0)=3\omega_q(\tau)=3\omega/2$, while $\omega\tau=30$ for (a) and $\omega\tau=40$ for (b). The phases are $\phi\approx 3\pi/4$ and $\phi\approx \pi$ for $\ket{\psi_{0,2}}$ and $\ket{\psi_{0,6}}$, respectively.  See Sec. V for further details. }}
\label{figCatextra}
\end{figure}

\section{Cat state preparation}\label{s:cat}
As explained in the main text, our method allows for the preparation of cat states of the bosonic mode.  In the main text we have illustrated with an example, denoted here by $\ket{\psi_{0,4}}=\frac{1}{\sqrt{2}}(\ket{0}+e^{i\phi}\ket{4})$ where $\ket{\psi_{n,m}}=\frac{1}{\sqrt{2}}(\ket{n}+e^{i\phi}\ket{m})$  denotes the cat state between $\ket{n}$ and $\ket{m}$, and where $\phi$ accounts for a relative phase, accumulated during the evolution. As considered in the main text, the STA evolution depends on the $n$-subspace of the JC Hamiltonian, and thus it is unable to complete exactly a transfer of population in two distinct subspaces at the same time. This unavoidably introduces errors in the cat state preparation, and thus the resulting fidelity is not exactly one, but close $F\approx 0.9992$ in the case considered in the main text, $\ket{\psi_{0,4}}$.

As further examples, we give here the values to obtain other cat states, such as  $\ket{\psi_{0,2}}$ and $\ket{\psi_{0,6}}$ (cf. Fig.~\ref{figCatextra}). Setting $\omega_q(0)=3\omega_q(\tau)=3\omega/2$ with $\lambda_0=0$ and $\lambda_m=\omega/4$ with $\omega\tau=30$, we find a fidelity  $F\approx 0.99998$ with $\phi\approx 3\pi/4$. The preparation of superpositions of further apart Fock states, such as $\ket{\psi_{0,6}}$, is more difficult due to the manipulation of different subspaces. However, one can still find high fidelities, $F\approx 0.988$ for $\omega\tau=40$ and $\ket{\psi_{0,6}}$ with $\phi\approx \pi$. It is worth stressing that depending on the specific target, the protocol can be adapted to obtain the desired state to a very good accuracy. In Fig.~\ref{figCatextra} we have plotted the resulting Wigner functions using the STA protocols, aiming to prepare (a) $\ket{\psi_{0,2}}$ and (b) $\ket{\psi_{0,6}}$. Furthermore, other states such as $\ket{\psi_{1,3}}$ (not shown explicitly), can be achieved with a similar fidelity, i.e. $F\approx 0.99$ for $\omega\tau=8$ and phase $\pi\approx \pi/20$, or $F\approx 0.9997$ for $\omega\tau\approx 3.2$ with  $\phi\approx 2\pi/5$.

In addition, we compare the example given in the main text ($\ket{\psi_{0,4}}$) for the time-dependent Jaynes-Cummings Hamiltonian without STA (bare) as well as for a time-independent (TI) and resonant $H_{\rm JC}$. For both the resulting fidelities are worse than the STA method, in particular $F\approx 0.91$ (bare) and $F\approx 0.70$ (TI). For $\ket{\psi_{0,2}}$  we find $F\approx 0.85$ (bare) and $F\approx 0.89$ (TI), while for $\ket{\psi_{0,6}}$ the resulting fidelities are $F\approx 0.69$ (bare) and $F\approx 0.32$ (TI). 
In the case of a time-independent $H_{\rm JC}$, we set $\omega_q=\omega$ and $\lambda={\rm max}_t \tilde{\lambda}(t)$, which for this case amounts to $\lambda=\lambda_m=\omega/4$. In this manner, an evolution during $t_{\pi,{\rm JC}}=\pi/(2 \lambda \sqrt{n+1})$ performs an effective $\pi$-pulse in the spin-boson dressed states with $n$ number of excitations. As the generation of the cat state requires population transfer in two different subspaces, the time-independent evolution is unable to correctly realize the cat state.

\section{Photon-shifted: thermal states and STA repetition}
\begin{figure}[t]
\centering
\includegraphics[width=1.\linewidth,angle=-0]{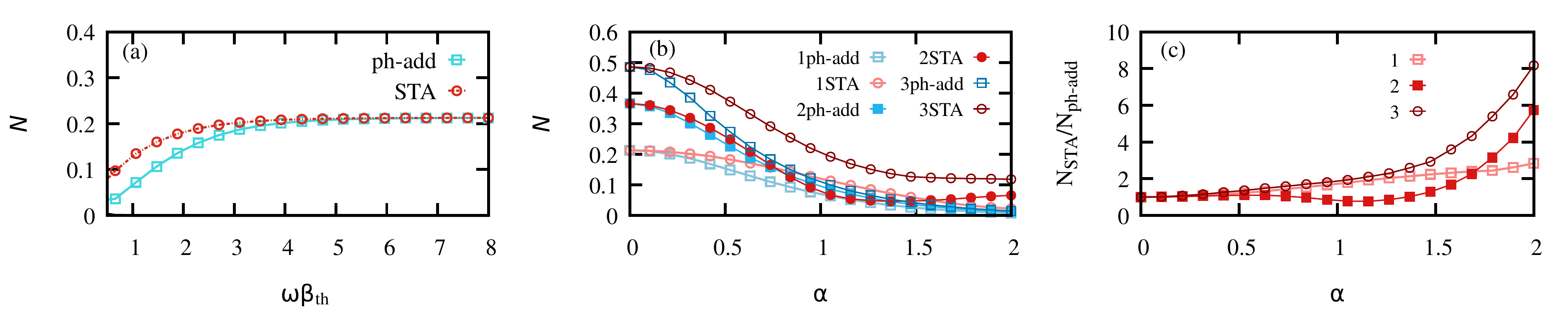}
\caption{\small{(a) Non-classicality of the state applying a STA evolution to remove the vacuum population of a thermal state at temperature $\beta_{\rm th}^{-1}$ (red dashed line) and its corresponding value by photon addition. In (b) we show the non-classicality $\mathcal{N}$ for an initial coherent state $\ket{e,\alpha}=D(\alpha)\ket{e,0}$ for both, STA and photon-addition, when performing them $n=1$, $2$ and $3$ times, as indicated in the legend. The ratio $\mathcal{N}_{\rm STA}/\mathcal{N}_{\rm ph-add}$ for the results shown in (b) is plotted in (c). See text for details about the parameters.}}
\label{figNonClass}
\end{figure}

As commented in the main text, our method allows for the generation of non-classical states when, for example, the vacuum population is transferred to the $\ket{1}$ Fock state. The achieved states are feature more non-classicality than those obtained by adding a photon, $\ket{\psi_{\rm ph-add}}\propto a^{\dagger}\ket{\psi}$. In the main text we have shown the non-classicality when applying it to an initial coherent state. Here, we show that this is also the case for thermal states, $\rho_{\rm th}=\frac{e^{-\beta_{\rm th}\omega\adaga}}{Z}\ket{n}\bra{n}$ with $Z={\rm Tr}[e^{-\beta_{\rm th}\omega\adaga}]$ and $\beta_{\rm th}$ denotes the inverse of the temperature. Recall that we quantify the non-classicality by integrating the region in which the Wigner function is negative~\cite{SMKenfack:04}
\begin{align}
  \mathcal{N}=\frac{1}{2\pi}\int d^2\beta (|W(\beta,\beta^*)|-W(\beta,\beta^*)),
\end{align}
where the Wigner function of a state $\rho$ is defined as $W(\beta,\beta^*)=2{\rm Tr}[\rho D(\beta)e^{i\pi\adaga}D^{\dagger}(\beta)]$~\cite{SMDavidovich:99}. As plotted in Fig.~\ref{figNonClass}(a), for very small temperatures, the STA method gives the same result as a photon addition to the thermal state as it approximately corresponds to the vacuum. However, as the temperature raises, photon addition yields less non-classicality than our method, in the a similar manner as for coherent states. As in previous cases, we consider $\omega\tau=8$, $\omega_q(0)=3\omega_q(\tau)=3\omega/2$ with $\lambda_0=0$ and $\lambda_m=\omega/4$. 

Remarkably, one can repeat this operation $n$ times to remove the populations in the Fock states $\ket{0}$, $\ket{1}$, $\ldots$, $\ket{n-1}$. As for the Fock state preparation, after each STA evolution, one needs to apply a $\pi$ pulse. We compute the non-classicality $\mathcal{N}$ of the states obtained when the protocol is performed $n=1$, $2$ and $3$ times and compare it with that of the $n$-photon added states, $\ket{\psi_{\rm n-ph-add}}\propto a^{\dagger,n}\ket{\psi}$. This is plotted in Fig.~\ref{figNonClass}(b) for an initial coherent state and same parameters as for panel (a). To better illustrate the enhancement with respect to simple photon addition, we show the ratio $\mathcal{N}_{\rm STA}/\mathcal{N}_{\rm ph-add}$ in Fig.~\ref{figNonClass}(c). Note that non-classicality of the state obtained using a STA protocol is larger than that of the photon-added state, i.e. $\mathcal{N}_{\rm STA}\geq N_{\rm ph-add}$, except for initial coherent states with $1\lesssim \alpha\lesssim 3/2$ upon $n=2$ STA evolutions, for which $\mathcal{N}_{\rm STA}$ becomes drops slightly below $\mathcal{N}_{\rm ph-add}$. 

\section{Robustness to imperfect pulse profiles}
\begin{figure}[t]
\centering
\includegraphics[width=1.\linewidth,angle=-0]{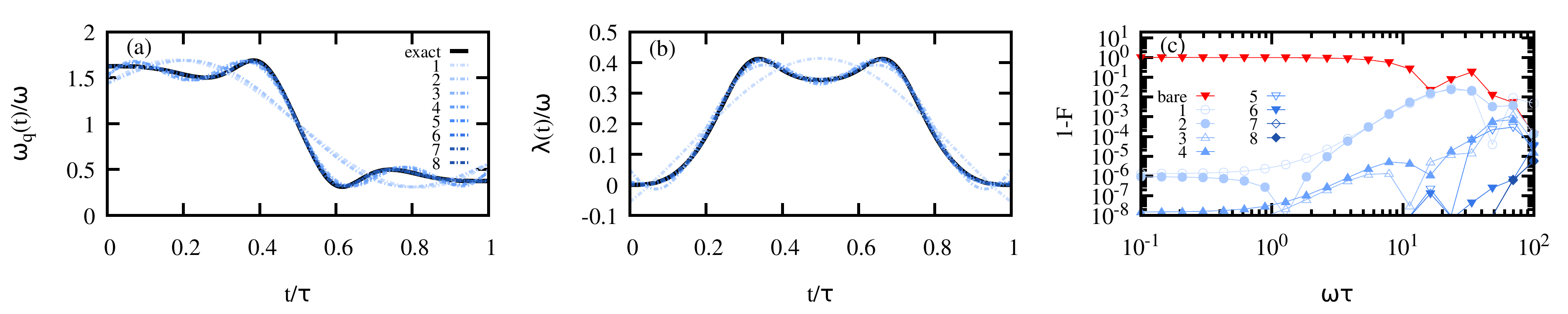}
\caption{\small{Profile of the time-dependent parameters, $\tilde{\omega}_q(t)$ and $\tilde{\lambda}(t)$ in (a) and (b), respectively, plotted with solid (black) lines for $\omega\tau=8$, $\omega_q(0)=3\omega_q(\tau)=3\omega/2$ with $\lambda_0=0$ and $\lambda_m=\omega/4$, and for the $n=0$ subspace. Dashed lines correspond to the approximated pulses using $N_{\rm F}$ Fourier modes, as indicated in legend. In (c) we show the infidelity $1-F$ with $F=|\langle g,1|\psi(\tau)\rangle|^2$ where $\ket{\psi(\tau)}$ is the final state after an evolution from $\ket{e,0}$ to $\ket{g,1}$ using the approximated pulses with $N_{\rm F}$ Fourier modes, and same parameters as before. For comparison, we plot the infidelity when using the bare Hamiltonian. Using the exact values of $\tilde{\omega}_q(t)$ and $\tilde{\lambda}(t)$ leads to $F=1$ by definition. }}
\label{figPulse}
\end{figure}

Eqs.~\eqref{eqS:wqtilde} and~\eqref{eqS:ltilde} provide the expressions for $\tilde{\omega}_q(t)$ and $\tilde{\lambda}(t)$, which dictate the STA evolution within the $n$-subspace of the JC model. Here we show the resilience of the performance to imperfect pulse implementation. For that, we approximate the actual form of the pulse $\tilde{\omega}_q(t)$ and $\tilde{\lambda}(t)$ with $N_{\rm F}$ Fourier modes, namely, via 
\begin{align}
\tilde{x}_{\rm F}(t)=\sum_{k=0}^{N_{\rm F}}c_k\cos(k\omega_{\rm F} t)+s_k\sin(k\omega_{\rm F} t)
\end{align}
with $\tilde{x}_{\rm F}(t)$ the denoting the approximation of $\tilde{x}(t)$ with $\tilde{x}\in\{\tilde{\omega}_q(t),\tilde\lambda\}$ using $N_{\rm F}$ modes. Note that we consider
\begin{align}\label{eqa:wqt}
  \omega_q(t)&=\omega_q(0)+10 \Delta\omega_q\, s^3 -15\Delta\omega_q s^4 +6\Delta\omega_qs^5,\\ \label{eqa:lt}
  \lambda(t)&=(\lambda_m-\lambda_0)\cos^4\left[\pi \left(1+2 s\right)/2 \right]+\lambda_0
\end{align}
with $s\!=\!t/\tau$, $\Delta\omega_q\!=\!\omega_q(\tau)-\omega_q(0)$. In Fig.~\ref{figPulse}(a) and (b) we show the approximated protocols for different $N_{\rm F}$ and for a particular case: $\omega\tau=8$, $\omega_q(0)=3\omega_q(\tau)=3\omega/2$ with $\lambda_0=0$ and $\lambda_m=\omega/4$, and for the $n=0$ subspace. Fig.~\ref{figPulse}(c) shows the infidelity with respect to the aimed state $\ket{g,1}$ when starting from $\ket{e,0}$ and using the previous approximated protocols. Increasing $N_{\rm F}$ the infidelity becomes arbitrarily small. Note that already for $N_{\rm F}\geq 3$ the state is retrieved with a very high fidelity $1-F<10^{-4}$. Moreover, even for the rough approximations of the pulses,  $N_{\rm F}=1$ and $2$, the infidelity becomes $1-F\approx 10^{-2}$  for an evolutions $\omega\tau\approx 10$, while $1-F$ reduces even more for shorter  evolutions. For comparison, we plot also the infidelity obtained under the bare Hamiltonian without implementing the STA protocols.

In addition, we investigate the robustness to pulse variations for the preparation of the two other non-classical states considered in the main text, namely, cat states (cf. Sec.~\ref{s:cat}) and photon-shifted states. Again, taking $N_{\rm F}$ from $1$ to $8$, we find that the cat state $\ket{\psi_{0,4}}$ is achieved with a high fidelity: $F>0.99$ already for $N_{\rm F}\geq 2$, while for $N_{\rm F}=1$ we observe $F=0.96$ ($N_{\rm F}=8$ leads to $F=0.9992$, essentially as the exact protocol). The parameters as the same as considered in the main text (Fig. 2(d)) and here in Sec.~\ref{s:cat}. We obtain similar results for different cat states, namely, for $\ket{\psi_{0,2}}$ and $\omega\tau=30$, we observe $F>0.99$ for $N_{\rm F}\geq 1$, while for $\ket{\psi_{0,6}}$ and $\omega\tau=40$ the fidelities are $F>0.98$ (close to the value under exact protocols, cf. Sec.~\ref{s:cat}) already for $N_{\rm F}\geq 4$. For photon-shifted states we observe similar results: for $\ket{e,\alpha}=D(\alpha)\ket{e,0}$ with $\alpha=3/4$ and $\omega\tau=8$ as chosen in the main text, we obtain $\mathcal{N}\approx 0.15$ for $N_{\rm F}\geq 1$, close to $\mathcal{N}= 0.1554$ under the exact protocols.

\begin{figure}[t]
\centering
\includegraphics[width=0.66\linewidth,angle=-0]{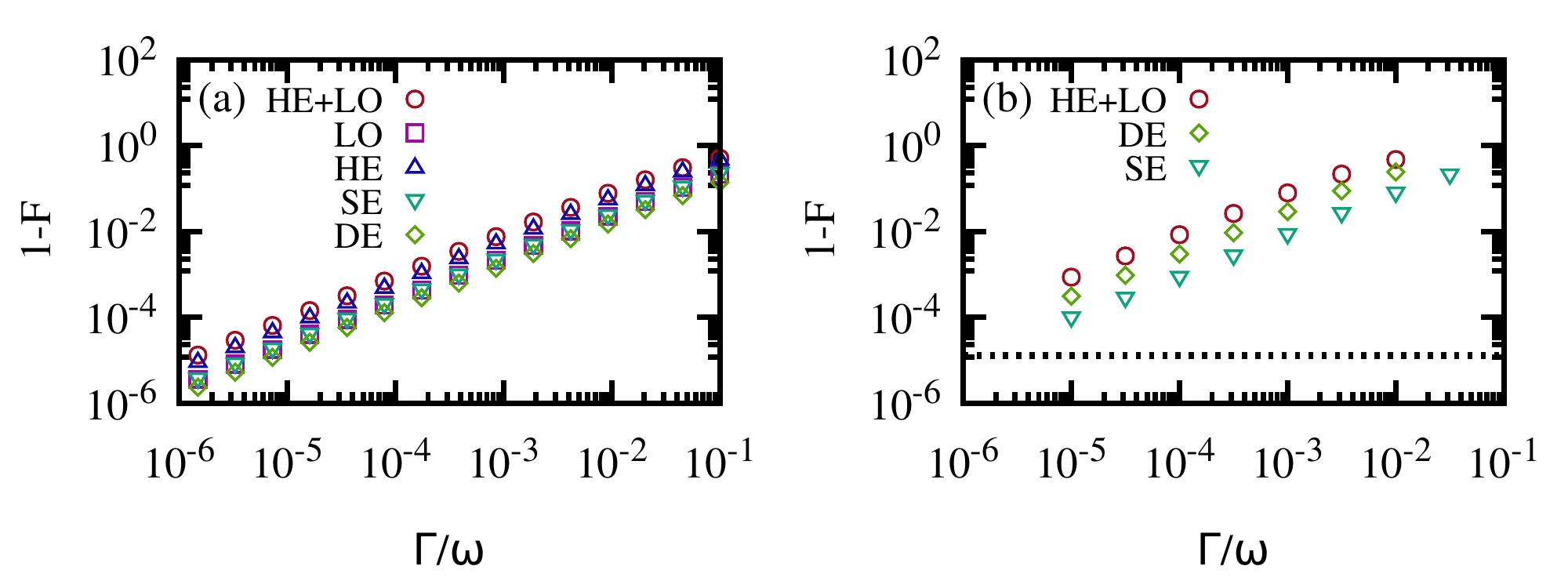}
\caption{\small{(a) Infidelity $1-F$ to achieve $\ket{g,1}$ from $\ket{e,0}$ using the STA protocol for $\omega\tau=5$, $\omega_q(0)=3\omega_q(\tau)=3\omega/2$ with $\lambda_0=0$ and $\lambda_m=\omega/4$ (as considered in Fig. 2(b) of main text), as a function of the noise rate $\Gamma/\omega$ for different decoherence processes, namely, heating plus losses $\Gamma_{a}=\Gamma_{\adag}$ (HE+LO), mode heating (HE) and losses (LO), spin spontaneous emission (SE) and spin dephasing (DE). In panel (b), we show the same analysis as in panel (a) but for the preparation of the cat state $\ket{\psi_{0,2}}=\frac{1}{\sqrt{2}}(\ket{0}+e^{i\phi}\ket{2})$. Dashed line shows the value of $1-F$ when $\Gamma=0$. For this case, the bare evolution and time-independent $H_{\rm JC}$ leads to $1-F\gtrsim 0.1$ regardless of $\Gamma$. }}
\label{figDecoh1}
\end{figure}

\section{Robustness to decoherence effects}
As commented in the main text, the presented method is robust against decoherence effects due to the ability to prepare non-classical states in a short time. For that we investigate the dynamics of the system when obeying the following master equation
\begin{align}\label{eqS:ME}
\dot{\rho}(t)=-i[H_x(t),\rho(t)]+\mathcal{D}_{\sigma^-}[\rho(t)]+\mathcal{D}_{\sigma_z}[\rho(t)]+\mathcal{D}_{a}[\rho(t)]+\mathcal{D}_{\adag}[\rho(t)]
  \end{align}
with $x\in\{{\rm JC},{\rm LCD}\}$ denoting the bare and STA protocols, respectively, and where  $\mathcal{D}_A[\bullet]=\Gamma_A(A\bullet A^\dagger-\{ A^\dagger A,\bullet\}/2)$ corresponds to the dissipator, in Lindblad form, of a jump operator $A$ with noise rate $\Gamma_A$. Note that we include spin spontaneous emission, spin dephasing, and absorption for the bosonic mode, as well as leakage or damping.

For that, we first consider the simplest case in which the state $\ket{e,0}$ is converted into $\ket{g,1}$ using the STA protocols. In Fig.~\ref{figDecoh1}(a) we show the impact of increasing the noise rate $\Gamma$ for different decoherence processes, namely, mode heating (HE) and losses (LO), spin dephasing (DE), and spin spontaneous emission (SE). The results for HE+LO are obtained assuming $\Gamma_{a}=\Gamma_{\adag}$.   It is worth stressing that the achieved fidelity under the bare Hamiltonian is extremely low, $F\approx 0.2$ for any $\Gamma$.

For the impact in cat state generation, we take first as example $\ket{\psi_{0,2}}=\frac{1}{\sqrt{2}}(\ket{0}+e^{i\phi}\ket{2})$. This is plotted in Fig.~\ref{figDecoh1}(b) using the same parameters as for Fig.~\ref{figCatextra}(a). We observe that heating and losses of the mode have a stronger impact than spin dephasing and spontaneous emission. In particular, for  $\Gamma_{a,\adag}/\omega\lesssim 10^{-3}$, one obtains $F\gtrsim 0.9$, while for $\Gamma_{\sigma_z}/\omega\lesssim 10^{-3}$ leads to $F\gtrsim 0.97$ and $\Gamma_{\sigma^-}/\omega\lesssim 10^{-3}$ leads to $F\gtrsim 0.99$. We emphasize however that these fidelities  may be improved by tuning system's parameters and searching for an optimal $\omega\tau$: while decreasing $\omega\tau$, the protocols show a stronger dependence on each of the distinct JC-subspaces, the impact of the decoherence processes is reduced.  Similar results are obtained for other cat states and photon-shifted states. For the latter we find that the negativity $\mathcal{N}\gtrsim 0.1$ when including all possible noises in the Eq.~\eqref{eqS:ME} with rates $\Gamma/\omega\lesssim 10^{-2}$ and $\omega\tau=8$ as considered in Fig. 3 of the main text. For $\Gamma_{a}=\Gamma_{\adag}=\Gamma_{\sigma^-}=\Gamma_{\sigma_z}=5\cdot 10^{-3}\omega$ we find $\mathcal{N}=0.12$, close to the value for noiseless evolution $\mathcal{N}=0.15$.

% Fock state, |0> + |4>, and photon-shifted \alpha=3/4. 

%\bibliographystyle{apsrev4-1.bst}
%\bibliography{QSE.bib}


\end{document}